# Three–dimensional lattice Boltzmann models for solid–liquid phase change

Dong Li, Ya-Ling He[*]

Key Laboratory of Thermo-Fluid Science and Engineering of Ministry of Education, School of Energy and Power Engineering, Xi'an Jiaotong University, China

[*]Corresponding email: yalinghe@mail.xjtu.edu.cn

**Abstract**

A three–dimensional (3 D) multiple–relaxation–time (MRT) and a 3 D single–relaxation–time (SRT) lattice Boltzmann (LB) models are proposed for the solid–liquid phase change. The enthalpy conservation equation can be recovered from the present models. The reasonable relationship of the relaxation times in the MRT model is discussed. Both One–dimensional (1 D) melting and solidification with analytical solutions are respectively calculated by the SRT and MRT models for validation. Compared with the SRT model, the MRT one is more accurate to capture the phase interface. The MRT model is also verified with other published two–dimensional (2 D) numerical results. The validations suggest that the present MRT approach is qualified to simulate the 3 D solid–liquid phase change process. Furthermore, the influences of Rayleigh number and Prandtl number on the 3 D melting are investigated.

**Key words**

Lattice Boltzmann method, Three–dimensional solid–liquid phase change, Enthalpy method, Natural convection

## 1. Introduction

The solid–liquid phase change is widely involved in the fields of latent heat storage used in the industrial waste heat recovery system [1], the building energy saving [2], and the solar power system [3, 4], the solidification processing in metals [5] and alloys [6], crystallization of crystals [7], and so on. Therefore, the research on the solid–liquid phase change has never been hung. In contrast with experiment, numerical methods are low–cost in both time and money [8]. The LB method might have a word to say in this context.

The LB method, chronologically first appearing in 1988 [9-11], has grown into a versatile, influential, and thriving numerical tool after almost three decades' development on both fundamental work and engineering applications [12-27]. The LB algorithm, evolving from the lattice gas automata (LGA), is a particle–based method [28]. The Boolean algebra is used for the operation of the particle numbers distributed by the Boolean fields in the LGA. For the purpose of statistical noise elimination, the local particle real distribution function is used to describe the particle contour, and the Boolean operation is substituted as real calculation [9, 28]. After four years' work [12-14], the LB method overcame all the drawbacks of its origin, the LGA, and became a hot topic. The LB method can be regarded as the discrete solver for the Boltzmann equation. Benefiting from the background of kinetic theory, the LB model can also be considered as a numerical method where the hydrodynamic behaviour on the macroscopic scale can be recovered on the basis of the microscopic kinetic principles [29]. The hydrodynamic parameters are calculated linearly from the particle distribution function. However, in the conventional computational fluid dynamics (CFD) methods, which focus on solving the discrete Navier–Stokes (N–S) equation, the convection term $\mathbf{u}\cdot\nabla\mathbf{u}$ is non–linear and the calculation of the pressure in the N–S equation should draw support from extra equation (such as the Poisson equation for the incompressible flow) [30].

Impressive achievements of the LB method can be found in major fields of classical fluid dynamics:

single–phase flows and heat transfer [24], flows and heat transfer in porous media [31, 32], multiphase flow [15, 26, 33-35], multi–component flow [36], turbulence [37], and multiple scale flow and heat transfer [38]. Furthermore, the burgeoning growth of the LB method in the field of solid–liquid phase change is also widely witnessed. The LB method was first extended to the solid–liquid phase transitions by Fabritiis and Succi et al. [39]. They drew an analogy between phase transitions and the chemical reaction, and used an LB model with chemical reactions to simulate solid–liquid phase change. Subsequently, Miller and Succi [40, 41] developed a solid–liquid phase–field LB method to simulate the anisotropic crystal growth from melting and melting flow of gallium. The interface–capturing equations, which are different in the liquid phase and the solid one, is used to trace the phase interface. Then, Miller et al. [42, 43] applied their model to explore the binary–alloy solidification as well as the microstructure evolving during the Czochralski growth of silicon–germanium. Semma and Mohamad et al. [44] used the temperature–based double–distribution–function (DDF) model to solve fluid flow and heat transfer with solid–liquid phase change. The phase interface was captured by using partial or probabilistic bounce back scheme. The enthalpy method [45] was first introduced into the LB approach by Jiaung et al. [46] to calculate the heat conduction with phase change. In Jiaung's enthalpy–updating scheme, the transient term $\partial h/\partial t$ in the enthalpy conservation equation without convection is divided into $\partial(C_pT)/\partial t$ and $\partial(f_lL)/\partial t$. Therefore, the enthalpy conservation equation is transformed into the temperature diffusion equation with a source term $\partial(f_lL/C_p)/\partial t$, which is solved by the LB model for diffusion. Chatterjee and Chakraborty [47] developed Jiaung's 2 D model to 3 D. Afterwards, they proposed a hybrid method for solid–liquid, where convection was considered [48, 49]. In their model, the enthalpy method was implement by finite difference method while the fluid flow in the liquid phase was calculated by using SRT LB model. Then, Huber et al. [50] as well as Chatterjee and Chakraborty [51] extended Jiaung's 2

D model, taking convection into account, respectively. Gao and Chen [52] developed a 2 D SRT DDF model to simulate the solid–liquid phase change with convection in porous media at the representative elementary volume (REV) scale. The local thermal equilibrium (LTE) between the metal matrix and the phase change material (PCM) is used in their model. Liu and He [53] proposed a double 2 D MRT model for solid–liquid phase change in porous media at REV scale, where LTE and convection were considered. Huang and Wu et al. [54] developed a new 2 D SRT model for solid–liquid phase change. The enthalpy method is accepted in Huang's model. Unlike the handling of the enthalpy method in Jiaung's scheme, the transient term $\partial h/\partial t$ in the enthalpy conservation equation is kept as one term rather than being divided. The enthalpy is directly solved by the SRT LB model. Huang and Wu further [55] promoted their 2 D SRT model as a 2 D MRT one. The superiority of the MRT model over the SRT one was discussed in their paper. Li and He et al. [56] proposed a enthalpy–based MRT model for axisymmetric solid–liquid phase change. Ren and Chan [57] explored the GPU acceleration of Huang and Wu's 2 D enthalpy–based MRT model. Liu and He [58] developed a 2 D enthalpy–based MRT model for solid–liquid phase change in metal foams, where the LTE condition was substituted as the more reasonable local thermal non-equilibrium (LTNE) condition.

Though much work has been devoted to the development of the LB model for solid–liquid phase change, the dimensions of the reported LB models are two. In this paper, 3 D enthalpy–based SRT and MRT models for solid–liquid phase change are both proposed. The rest of the paper is organized as follows: In Section 2, the 3 D enthalpy–based SRT and MRT models will be introduced elaborately, respectively. The D3Q19 MRT model for fluid flow in the liquid phase domain reported in Ref. [20] are described in brief. In Section 3, the present models are compared with not only analytical solutions but also the published numerical results and experimental data for validation. The paper is concluded following a discussion in Section 4.

## 2. Model description

2.1 The enthalpy conservation equation with incompressible flow

The energy conservation equation with incompressible flow can be written as:

$$\frac{\partial(h)}{\partial t} + \frac{\partial(C_p T u_i)}{\partial x_i} = \frac{\partial}{\partial x_i}\left(\frac{1}{\rho_0}\frac{\partial(\lambda \nabla T)}{\partial x_i}\right) \tag{1}$$

where $t$ is the time, $x_i$ is the Cartesian spatial coordinate, $C_p$ is the specific heat at constant pressure, $T$ is the temperature, $u_i$ is the velocity, and $\lambda$ is the thermal conductivity.

In addition, the enthalpy $h$ can be expressed as

$$h = C_p T + f_l L - h_0 \tag{2}$$

where $f_l$ is the liquid phase volume fraction, $L$ is the latent heat of phase change, and the reference enthalpy $h_0$ equals to 0 J kg$^{-1}$ at 0 K.

Eq. (2) can be recovered from the LB models proposed in this paper via the Chapman–Enskog expanding, which means that our models are alternative numerical methods to solve Eq. (2).

2.2 SRT model for 3 D solid–liquid phase change

The kernel of the LB model is the evolution of the microscopic particle distribution function. The microscopic particle velocities are discretized to seven lattice velocities for the three–dimensional solid–liquid phase change LB model in this paper. The seven lattice velocities are defined as:

$$[\mathbf{e}_0 \ \mathbf{e}_1 \ \mathbf{e}_2 \ \mathbf{e}_3 \ \mathbf{e}_4 \ \mathbf{e}_5 \ \mathbf{e}_6] = \begin{bmatrix} 0 & 1 & -1 & 0 & 0 & 0 & 0 \\ 0 & 0 & 0 & 1 & -1 & 0 & 0 \\ 0 & 0 & 0 & 0 & 0 & 1 & -1 \end{bmatrix} c \tag{3}$$

where $c = \delta_x/\delta_t$ is the lattice constant.

For D3Q7 (3 dimension 7 discrete lattice velocities), the lattice velocities satisfied the relations as follows:

$$\sum_{\alpha=0}^{6} e_{\alpha i} = 0 \qquad \sum_{\alpha=0}^{6} e_{\alpha i} e_{\alpha j} = 2c^2 \delta_{ij} \qquad (4)$$

The evolution equation in the velocity space spanned by $\mathbf{e}_\alpha$ is:

$$|g(\mathbf{x}+\mathbf{e}_\alpha \delta_t, t+\delta_t)\rangle = |g(\mathbf{x},t)\rangle + \Omega|g(\mathbf{x},t)\rangle \qquad (5)$$

where $\Omega|g(\mathbf{x},t)\rangle$ is the collision operator.

According to the Bhatnagar–Gross–Krook (BGK) approximation, the collision operator for the SRT model is [14]:

$$\Omega|g(\mathbf{x},t)\rangle = -\frac{1}{\tau_g}\left(|g(\mathbf{x},t)\rangle - |g^{eq}(\mathbf{x},t)\rangle\right) \qquad (6)$$

The relaxation times for all the components of the distribution function are uniform in the SRT model.

The equilibrium distribution function is defined as:

$$g_\alpha^{eq}(\mathbf{x},t) = \begin{cases} h - (1-\omega_\alpha) C_{p,\text{ref}} T & \alpha = 0 \\ \omega_\alpha C_{p,\text{ref}} T + \dfrac{\mathbf{u}\cdot\mathbf{e}_\alpha}{c_{sT}^2} \omega_\alpha C_p T & \alpha \neq 0 \end{cases} \qquad (7)$$

where $C_{p,\text{ref}}$ is the reference specific heat at constant pressure, and the lattice sound speed $c_{sT}$ equals to $(\omega_T)^{1/2} c$. $\omega_T$ is a constant.

In addition, the weight factor $\omega_\alpha$ is given as:

$$\omega_\alpha = \begin{cases} 1 - 3\omega_T & \alpha = 0 \\ \dfrac{1}{2}\omega_T & \alpha \neq 0 \end{cases} \qquad (8)$$

Calculating the zeroth–, first– and second–order moment of the equilibrium distribution function respectively, we obtain:

$$\sum_{\alpha=0}^{6} g_\alpha^{eq} = h, \quad \sum_{\alpha=0}^{6} e_{\alpha i} g_\alpha^{eq} = C_p T u_i, \quad \sum_{\alpha=0}^{6} e_{\alpha i} e_{\alpha j} g_\alpha^{eq} = C_{p,\text{ref}} T \omega_T c^2 \delta_{ij} \qquad (9)$$

After the Taylor series expansion, the evolution equation can be written as:

$$\left(\frac{\partial}{\partial t} + \frac{\partial}{\partial x} e_{\alpha i}\right) \delta_t g_\alpha(\mathbf{x},t) + \frac{1}{2}\left(\frac{\partial}{\partial t} + \frac{\partial}{\partial x} e_{\alpha i}\right)^2 \delta_t^2 g_\alpha(\mathbf{x},t) = -\frac{1}{\tau_g}\left(g(\mathbf{x},t) - g^{eq}(\mathbf{x},t)\right) + O(\delta_t^3) \qquad (10)$$

The distribution function $g_\alpha$ and the derivatives of the time can be expanded as follows:

$$g_\alpha = g_\alpha^{eq} + \delta_t g_\alpha^{(1)} + \delta_t^2 g_\alpha^{(2)}, \quad \frac{\partial}{\partial t} = \frac{\partial}{\partial t_0} + \delta_t \frac{\partial}{\partial t_1} \tag{11}$$

Substituting Eq. (11) into Eq. (10), we can rewrite Eq. (10) on the $\delta_t$ and $\delta_t^2$ scales respectively:

$$\delta_t: \quad \left(\frac{\partial}{\partial t_0} + e_{\alpha i}\frac{\partial}{\partial x_i}\right) g_\alpha^{eq} = -\frac{1}{\tau_g} g_\alpha^{(1)} \tag{12}$$

$$\delta_t^2: \quad \frac{\partial g_\alpha^{eq}}{\partial t_1} + \left(\frac{\partial}{\partial t_0} + e_{\alpha i}\frac{\partial}{\partial x_i}\right) g_\alpha^{(1)} + \frac{1}{2}\left(\frac{\partial}{\partial t_0} + e_{\alpha i}\frac{\partial}{\partial x_i}\right)^2 g_\alpha^{eq} = -\frac{1}{\tau_g} g_\alpha^{(2)} \tag{13}$$

substitute Eq. (12) into Eq. (13):

$$\delta_t^2: \quad \frac{\partial g_\alpha^{eq}}{\partial t_1} + \left(1 - \frac{1}{2\tau_g}\right)\left(\frac{\partial}{\partial t_0} + e_{\alpha i}\frac{\partial}{\partial x_i}\right) g_\alpha^{(1)} = -\frac{1}{\tau_g} g_\alpha^{(2)} \tag{14}$$

Calculating the zeroth–order moment of Eq. (10) on the $\delta_t$ scale, we obtain:

$$\delta_t: \quad \frac{\partial h}{\partial t_0} + \frac{\partial}{\partial x_i}\left(C_p T u_i\right) = 0 \tag{15}$$

Analogously, the zeroth–order moment of Eq. (10) on the $\delta_t^2$ scale is:

$$\delta_t^2: \quad \frac{\partial h}{\partial t_1} + \left(1 - \frac{1}{2\tau_g}\right)\frac{\partial}{\partial x_i}\left(\sum_\alpha e_{\alpha i} g_\alpha^{(1)}\right) = 0 \tag{16}$$

Using Eq. (9) and (12), and neglecting the high order term, we can derive the first–order moment of $g_\alpha^{(1)}$:

$$\sum_\alpha e_{\alpha i} g_\alpha^{(1)} = -\tau_g \frac{\partial}{\partial x_i}\left(C_{p,\text{ref}} T \omega_T c^2\right) \tag{17}$$

Substituting Eq. (17) into Eq. (16), we can rewrite Eq. (16):

$$\delta_t^2: \quad \frac{\partial h}{\partial t_1} + \frac{\partial}{\partial x_i}\left(\left(\frac{1}{2} - \tau_g\right)\omega_T c^2 \frac{\partial}{\partial x_i}\left(C_{p,\text{ref}} T\right)\right) = 0 \tag{18}$$

Combining Eq. (15) with Eq. (18), we can obtain the target macroscopic enthalpy conservation equation:

$$\frac{\partial h}{\partial t} + \frac{\partial}{\partial x_i}\left(C_p T u_i\right) = \frac{\partial}{\partial x_i}\left(\chi \frac{\partial}{\partial x_i}\left(C_{p,\text{ref}} T\right)\right) \tag{19}$$

where the thermal diffusivity $\chi$ equals to:

$$\chi = \left(\tau_g - \frac{1}{2}\right)\omega_T c^2 \delta_t \tag{20}$$

2.3 MRT model for 3 D solid–liquid phase change

The SRT model is popular in the LB community owing to its simplicity. However, it suffers from some obvious shortcomings. For instance, the SRT model is severely numerical instable when the relaxation time is close to 0.5. The MRT model has attracted much attention since 2000 [19, 20], which have a better numerical stability than the SRT model. In the MRT model, the streaming processes of distribution functions are also carried out in the velocity moment spanned by the discrete lattice velocities shown in Eq. (3). However, in contrast to the SRT model, the collision of each distribution function with respective relaxation time is completed in the moment space in the MRT model, which can be written as:

$$|n^+\rangle = |n\rangle - \mathbf{R}(|n\rangle - |n^{eq}\rangle) \tag{21}$$

where the distribution function and the relaxation matrix in the moment space can be expressed respectively as:

$$|n\rangle = \mathbf{N}[g_0, g_1, ..., g_6]^T \tag{22}$$

$$\mathbf{R} = \mathrm{diag}(\sigma_0, \sigma_1, \sigma_2, \sigma_3, \sigma_4, \sigma_5, \sigma_6) \tag{23}$$

where $\sigma_1 = \sigma_2 = \sigma_3 = \sigma_\chi$.

The transformation matrix $\mathbf{N}$, which connects the velocity space with the moment space, can be expressed as [59]:

$$\mathbf{N} = \begin{bmatrix} 1 & 1 & 1 & 1 & 1 & 1 & 1 \\ 0 & 1 & -1 & 0 & 0 & 0 & 0 \\ 0 & 0 & 0 & 1 & -1 & 0 & 0 \\ 0 & 0 & 0 & 0 & 0 & 1 & -1 \\ 6 & -1 & -1 & -1 & -1 & -1 & -1 \\ 0 & 2 & 2 & -1 & -1 & -1 & -1 \\ 0 & 0 & 0 & 1 & 1 & -1 & -1 \end{bmatrix} \tag{24}$$

The equilibrium distribution function in the moment space is defined as:

$$|n^{eq}\rangle = \left(h, \frac{C_p T u_x}{c}, \frac{C_p T u_y}{c}, \frac{C_p T u_z}{c}, 6h - 21\omega_T C_{p,\mathrm{ref}} T, 0, 0\right)^T \tag{25}$$

The evolution equation in the MRT model, to which applied the Taylor series expansion, can be rewritten as:

$$(\partial_t \mathbf{I} + \mathbf{E}_i \partial_i)\mathbf{n} + \frac{\delta_t}{2}(\partial_t \mathbf{I} + \mathbf{E}_i \partial_i)^2 \mathbf{n} + \frac{\mathbf{R}}{\delta_t}(\mathbf{n} - \mathbf{n}^{eq}) + O(\delta_t^2) = 0 \qquad (26)$$

where $\mathbf{E}_\alpha = \mathbf{M}\mathbf{e}_\alpha \mathbf{M}^{-1}$.

As the beginning of the Chapman–Enskog expansion, $\mathbf{n}$ is expended as:

$$\mathbf{n} = \mathbf{n}^{eq} + \delta_t \mathbf{n}^{(1)} + \delta_t^2 \mathbf{n}^{(2)} + \ldots \qquad (27)$$

Substituting Eq. (27) into Eq. (26), after some algebraic operation, we obtain:

$$\delta_t: \qquad (\partial_t \mathbf{I} + \mathbf{E}_i \partial_i)\mathbf{n}^{eq} + \mathbf{R}\mathbf{n}^{(1)} = 0 \qquad (28)$$

$$\delta_t^2: \qquad \partial_{t_1} \mathbf{n}^{eq} + (\mathbf{I} - 0.5\mathbf{R})\partial_{t_0}\mathbf{n}^{(1)} + \mathbf{E}_i \partial_i (\mathbf{I} - 0.5\mathbf{R})\mathbf{n}^{(1)} + \mathbf{R}\mathbf{n}^{(2)} = 0 \qquad (29)$$

The conserved moment $n_0$ satisfies $n_0^{(1)} = n_0^{(2)} = \ldots = 0$.

With the help of Eq. (26), the equations connected with $n_0$ in Eq. (28) and Eq. (29) can be written as:

$$\frac{\partial h}{\partial t_0} + \frac{\partial}{\partial x_i}(u_i C_p T) = 0 \qquad (30)$$

$$\frac{\partial h}{\partial t_1} + c\nabla \cdot \begin{pmatrix} (1 - 0.5\sigma_1)n_1^{(1)} \\ (1 - 0.5\sigma_2)n_2^{(1)} \\ (1 - 0.5\sigma_3)n_3^{(1)} \end{pmatrix} = 0 \qquad (31)$$

Using Eq. (28), we rewrite Eq. (31):

$$\frac{\partial h}{\partial t_1} = \frac{\partial}{\partial x_i}\left(\left(\frac{1}{\sigma_\chi} - \frac{1}{2}\right)\omega_T c^2 \delta_t \frac{\partial}{\partial x_i}(C_{p,\text{ref}} T)\right) \qquad (32)$$

With combining Eq. (30) and Eq. (32), the macroscopic enthalpy conservation equation can be recovered as:

$$\frac{\partial h}{\partial t} + \frac{\partial}{\partial x_i}(C_p T u_i) = \frac{\partial}{\partial x_i}\left(\chi \frac{\partial}{\partial x_i}(C_{p,\text{ref}} T)\right) \qquad (33)$$

where the thermal diffusivity $\chi$ is defined as:

$$\chi = \left(\frac{1}{s_\chi} - \frac{1}{2}\right)\omega_T c^2 \delta_t \qquad (34)$$

2.4 The relationship of the relaxation times in the MRT model

The seven relaxation times on the main diagonal of the relaxation matrix make the MRT model flexible in application. However, if the relaxation times are selected irrelevantly, numerical deviations will appear [60, 61]. Especially, the aforementioned circumstance will cause the numerical diffusion across the solid–liquid phase interface in the solid–liquid phase change simulation [55]. In order to eliminate this error, we will determine the relations between the relaxation times in the present MRT model via the analysis of melting case dominated by the conduction. For convenience, only the $z$–direction conduction is considered. The initial temperature equals the melting temperature. Five lattices are chosen in the analysis. As shown in Fig. 1, the middle lattice $i_0$ is the phase interface. The left–hand side (LHS) of the phase interface is liquid phase while the right–hand side (RHS) of the phase interface is solid phase at the melting temperature.

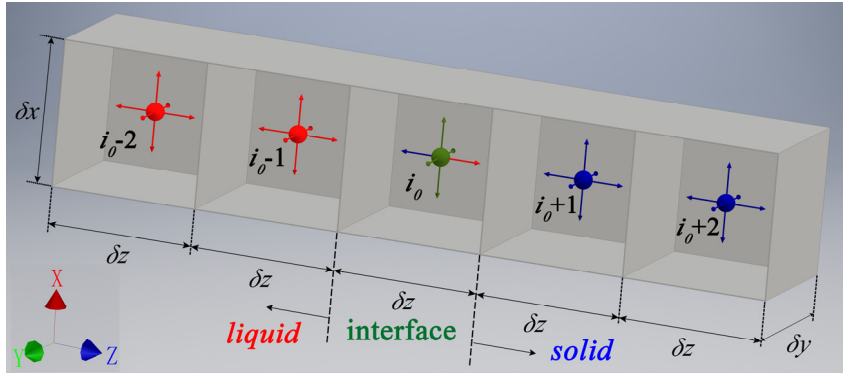

Fig. 1 Illustration of five lattices used in the relationship analysis of seven relaxation times

After the collision in the moment space, the distribution functions at space $i$ and time $t + 1$ in the velocity space can be expressed as:

$$g_\alpha^{i,t+1} = \begin{cases} g_\alpha^{i,t} - r_{\alpha k}^{i,t}\left(g_k^{i,t} - g_k^{eq,i,t}\right) & \alpha = 0,3,4,5,6 \\ g_\alpha^{i+1,t} - r_{\alpha k}^{i+1,t}\left(g_k^{i+1,t} - g_k^{eq,i+1,t}\right) & \alpha = 2 \\ g_\alpha^{i-1,t} - r_{\alpha k}^{i-1,t}\left(g_k^{i-1,t} - g_k^{eq,i-1,t}\right) & \alpha = 1 \end{cases} \quad (35)$$

where $r_{\alpha k}^{i,t} = \mathbf{M}^{-1}\mathbf{R}\mathbf{M}$.

After some algebraic manipulations, we can deduce three equations form Eq. (35) and Eq. (9):

$$g_0^{i,t+1} + g_3^{i,t+1} + g_4^{i,t+1} + g_5^{i,t+1} + g_6^{i,t+1} = (1-\sigma_5)\left(g_0^{i,t} + g_3^{i,t} + g_4^{i,t} + g_5^{i,t} + g_6^{i,t}\right) \tag{36}$$
$$+ \sigma_5 \left(h^{i,t} - \omega_T C_{p,\text{ref}} T^{i,t}\right)$$

$$g_2^{i,t+1} = (1-\sigma_1^{i+1,t}) g_2^{i+1,t} + \frac{\sigma_4^{i+1,t} - \sigma_1^{i+1,t}}{2}\left(g_0^{i+1,t} + g_3^{i+1,t} + g_4^{i+1,t} + g_5^{i+1,t} + g_6^{i+1,t}\right)$$
$$- \frac{\sigma_1^{i+1,t} - \sigma_4^{i+1,t}}{2}\left(h^{i+1,t} - \omega_T C_{p,\text{ref}} T^{i+1,t}\right) + \sigma_1^{i+1,t}\left(\frac{1}{2}\omega_T C_{p,\text{ref}} T^{i+1,t}\right) \tag{37}$$

$$g_1^{i,t+1} = (1-\sigma_1^{i-1,t}) g_1^{i-1,t} + \frac{\sigma_4^{i-1,t} - \sigma_1^{i-1,t}}{2}\left(g_0^{i-1,t} + g_3^{i-1,t} + g_4^{i-1,t} + g_5^{i-1,t} + g_6^{i-1,t}\right)$$
$$- \frac{\sigma_4^{i-1,t} - \sigma_1^{i-1,t}}{2}\left(h^{i-1,t} - \omega_T C_{p,\text{ref}} T^{i-1,t}\right) + \sigma_1^{i-1,t}\left(\frac{1}{2}\omega_T C_{p,\text{ref}} T^{i-1,t}\right) \tag{38}$$

In the solid phase ($i \geq i_0 + 1$), $T \equiv T_m$, $h \equiv h_m = C_{p,s} T_m$. Then, Eq. (36) can be rewritten as:

$$g_0^{i,t+1} + g_3^{i,t+1} + g_4^{i,t+1} + g_5^{i,t+1} + g_6^{i,t+1} - (h_m - \omega_T C_{p,\text{ref}} T_m) = (1-\sigma_5)\left(g_0^{i,t} + g_3^{i,t} + g_4^{i,t} + g_5^{i,t} + g_6^{i,t}\right. \tag{39}$$
$$\left. - (h_m - \omega_T C_{p,\text{ref}} T_m)\right)$$

Owing to $0 < \sigma_5 < 2$, namely $-1 < 1 - \sigma_5 < 1$, Eq. (39) is satisfied when:

$$g_0^{i,t+1} + g_3^{i,t+1} + g_4^{i,t+1} + g_5^{i,t+1} + g_6^{i,t+1} \equiv (h_m - \omega_T C_{p,\text{ref}} T_m) \tag{40}$$

$T$ equals $T_m$ at the phase interface and in the solid phase ($i \geq i_0$). Using Eq. (40), we can reduce Eq. (37) to:

$$g_2^{i,t+1} - \frac{1}{2}\omega_T C_{p,\text{ref}} T_m = (1-\sigma_1^{i+1,t})\left(g_2^{i+1,t} - \frac{1}{2}\omega_T C_{p,\text{ref}} T_m\right) \tag{41}$$

Similarly, $-1 < 1 - \sigma_1 < 1$, the solution of Eq. (41) is:

$$g_2^{i,t+1} = \frac{1}{2}\omega_T C_{p,\text{ref}} T_m \tag{42}$$

$T^{i_0+1}$ equals $T_m$ and $h^{i_0+1}$ equals $h_m$ for $i \geq i_0 + 2$. Substituting Eq. (40) into Eq. (38), we obtain:

$$g_1^{i,t+1} - \frac{1}{2}\omega_T C_{p,\text{ref}} T_m = (1-\sigma_1^{i-1,t})\left(g_1^{i-1,t} - \frac{1}{2}\omega_T C_{p,\text{ref}} T_m\right) \tag{43}$$

According to Eq. (43), if $g_1^{i-1,t+1}$ is not equal to $0.5\omega_T C_{p,\text{ref}} T_m$, the error term (the RES of Eq. (43)) will transfer from $i_0 + 1$ into $i_0 + 2$, which makes the enthalpy at $i_0 + 2$ equals $h_m$ + the error term rather than $h_m$. In order to eliminate the deviation, we now focus on the lattice $i_0 + 1$. Using Eq. (9) and Eq. (7), we can deduce $g_1$ at $i_0$:

$$g_1^{i_0,t} = h^{i_0,t} - \left(g_0^{i_0,t} + g_3^{i_0,t} + g_4^{i_0,t} + g_5^{i_0,t} + g_6^{i_0,t}\right) - \frac{1}{2}\omega_T C_{p,\text{ref}} T_m \tag{44}$$

After substituting Eq. (44) into Eq. (38) and some algebraic manipulations, we deduce:

$$g_1^{i_0+1,t+1} - \frac{1}{2}\omega_T C_{p,\text{ref}} T_m = \frac{\sigma_1^{i_0,t} + \sigma_4^{i_0,t} - 2}{2}\left(g_0^{i_0,t} + g_3^{i_0,t} + g_4^{i_0,t} + g_5^{i_0,t} + g_6^{i_0,t} - h^{i_0,t} + \omega_T C_{p,\text{ref}} T_m\right) \quad (45)$$

From Eq. (43) and Eq. (45), it can be seen that the error term at lattice $i_0 + 1$, the RHS of Eq. (45), come from the lattice $i_0$. Further, the error term at lattice $i_0 + 1$ transfers to lattice $i_0 + 2$. Then, the error term at lattice $i_0 + 2$ rises to the surface:

$$\left(1 - \sigma_1^{i_0+1,t}\right)\left(\frac{\sigma_1^{i_0,t} + \sigma_4^{i_0,t} - 2}{2}\right)\left(g_0^{i_0,t} + g_3^{i_0,t} + g_4^{i_0,t} + g_5^{i_0,t} + g_6^{i_0,t} - h^{i_0,t} + \omega_T C_{p,\text{ref}} T_m\right) \quad (46)$$

The source of the deviation at lattice $i_0 + 2$, clearly seen in Eq. (46), comes from lattice $i_0$ which is the phase interface. The error term will continue to spread from the phase interface to the depths of the solid phase. According to Eq. (45), the method to strangle the deviation at lattice $i_0 + 1$ is:

$$\sigma_1 + \sigma_4 = 2 \quad (47)$$

2.5 MRT model for fluid flow

The D3Q19 MRT model for fluid flow proposed by d'Humieres and Krafczyk et. al [20] is used in the liquid phase domain. The details about this model can be found in Ref. [20]. For the sake of readers' convenience, brief introduction will be made in this section.

The nineteen lattice velocities are defined as:

$$\mathbf{e} = \begin{bmatrix} 0 & 1 & -1 & 0 & 0 & 0 & 0 & 1 & -1 & 1 & -1 & 1 & -1 & 1 & -1 & 0 & 0 & 0 & 0 \\ 0 & 0 & 0 & 1 & -1 & 0 & 0 & 1 & 1 & -1 & -1 & 0 & 0 & 0 & 0 & 1 & -1 & 1 & -1 \\ 0 & 0 & 0 & 0 & 0 & 1 & -1 & 0 & 0 & 0 & 0 & 1 & 1 & -1 & -1 & 1 & 1 & -1 & -1 \end{bmatrix} \quad (48)$$

The evolution of the distribution function is divided into the collision in the moment space and the streaming in the velocity space, which are expressed as follows, respectively:

$$|m^+\rangle = |m\rangle - \mathbf{\Lambda}\left(|m\rangle - |m^{\text{eq}}\rangle\right) + \delta_t\left(\mathbf{I} - \frac{1}{2}\mathbf{\Lambda}\right)|FF\rangle \quad (49)$$

$$|f(\mathbf{x} + \mathbf{e}_\alpha \delta_t, t + \delta_t)\rangle = \mathbf{M}^{-1}|m^+(\mathbf{x}, t)\rangle \quad (50)$$

where the transformation matrix $\mathbf{M}$ can be written as:

$$\mathbf{M} = \begin{bmatrix}
1 & 1 & 1 & 1 & 1 & 1 & 1 & 1 & 1 & 1 & 1 & 1 & 1 & 1 & 1 & 1 & 1 & 1 & 1 \\
-30 & -11 & -11 & -11 & -11 & -11 & -11 & 8 & 8 & 8 & 8 & 8 & 8 & 8 & 8 & 8 & 8 & 8 & 8 \\
12 & -4 & -4 & -4 & -4 & -4 & -4 & 1 & 1 & 1 & 1 & 1 & 1 & 1 & 1 & 1 & 1 & 1 & 1 \\
0 & 1 & -1 & 0 & 0 & 0 & 0 & 1 & -1 & 1 & -1 & 1 & -1 & 1 & -1 & 0 & 0 & 0 & 0 \\
0 & -4 & 4 & 0 & 0 & 0 & 0 & 1 & -1 & 1 & -1 & 1 & -1 & 1 & -1 & 0 & 0 & 0 & 0 \\
0 & 0 & 0 & 1 & -1 & 0 & 0 & 1 & 1 & -1 & -1 & 0 & 0 & 0 & 0 & 1 & -1 & 1 & -1 \\
0 & 0 & 0 & -4 & 4 & 0 & 0 & 1 & 1 & -1 & -1 & 0 & 0 & 0 & 0 & 1 & -1 & 1 & -1 \\
0 & 0 & 0 & 0 & 0 & 1 & -1 & 0 & 0 & 0 & 0 & 1 & 1 & -1 & -1 & 1 & 1 & -1 & -1 \\
0 & 0 & 0 & 0 & 0 & -4 & 4 & 0 & 0 & 0 & 0 & 1 & 1 & -1 & -1 & 1 & 1 & -1 & -1 \\
0 & 2 & 2 & -1 & -1 & -1 & -1 & 1 & 1 & 1 & 1 & 1 & 1 & 1 & 1 & -2 & -2 & -2 & -2 \\
0 & -4 & -4 & 2 & 2 & 2 & 2 & 1 & 1 & 1 & 1 & 1 & 1 & 1 & 1 & -2 & -2 & -2 & -2 \\
0 & 0 & 0 & 1 & 1 & -1 & -1 & 1 & 1 & 1 & 1 & -1 & -1 & -1 & -1 & 0 & 0 & 0 & 0 \\
0 & 0 & 0 & -2 & -2 & 2 & 2 & 1 & 1 & 1 & 1 & -1 & -1 & -1 & -1 & 0 & 0 & 0 & 0 \\
0 & 0 & 0 & 0 & 0 & 0 & 0 & 1 & -1 & -1 & 1 & 0 & 0 & 0 & 0 & 0 & 0 & 0 & 0 \\
0 & 0 & 0 & 0 & 0 & 0 & 0 & 0 & 0 & 0 & 0 & 0 & 0 & 0 & 0 & 1 & -1 & -1 & 1 \\
0 & 0 & 0 & 0 & 0 & 0 & 0 & 0 & 0 & 0 & 0 & 1 & -1 & -1 & 1 & 0 & 0 & 0 & 0 \\
0 & 0 & 0 & 0 & 0 & 0 & 0 & 1 & -1 & 1 & -1 & -1 & 1 & -1 & 1 & 0 & 0 & 0 & 0 \\
0 & 0 & 0 & 0 & 0 & 0 & 0 & -1 & -1 & 1 & 1 & 0 & 0 & 0 & 0 & 1 & -1 & 1 & -1 \\
0 & 0 & 0 & 0 & 0 & 0 & 0 & 0 & 0 & 0 & 0 & 1 & 1 & -1 & -1 & -1 & -1 & 1 & 1
\end{bmatrix}$$

(51)

The relaxation matrix $\mathbf{\Lambda}$ is as follows:

$$\mathbf{\Lambda} = \text{diag}\left(0, s_e, s_\varepsilon, 0, s_q, 0, s_q, 0, s_q, s_\nu, s_\pi, s_\nu, s_\pi, s_\nu, s_\nu, s_\nu, s_t, s_t, s_t\right) \quad (52)$$

The equilibrium distribution function is defined as:

$$\left|m^{eq}\right\rangle = \rho\left(1, -11 + 19\frac{u^2}{c^2}, 3 - \frac{11u^2}{2c^2}, \frac{u_x}{c}, -\frac{2u_x}{3c}, \frac{u_y}{c}, -\frac{2u_y}{3c}, \frac{u_z}{c}, -\frac{2u_z}{3c}, \frac{3u_x^2 - u^2}{c^2}, \right.$$
$$\left. \frac{-u_x^2 + 0.5u_y^2 + u_z^2}{c^2}, \frac{u_y^2 - u_z^2}{c^2}, \frac{u_z^2 - u_y^2}{2c^2}, \frac{u_x u_y}{c^2}, \frac{u_y u_z}{c^2}, \frac{u_z u_x}{c^2}, 0, 0, 0\right)^{\text{T}} \quad (53)$$

The forcing term $|FF\rangle$ in the moment space can be written as:

$$|FF\rangle = \left(0, \frac{38\mathbf{F} \cdot \mathbf{u}}{c^2}, -\frac{11\mathbf{F} \cdot \mathbf{u}}{c^2}, \frac{F_x}{c}, -\frac{2F_x}{3c}, \frac{F_y}{c}, -\frac{2F_y}{3c}, \frac{F_z}{c}, -\frac{2F_z}{3c}, \frac{2F_x u_x - 2\mathbf{F} \cdot \mathbf{u}}{c^2}, \frac{\mathbf{F} \cdot \mathbf{u} - 3F_x u_x}{c^2}, \right.$$
$$\left. \frac{2F_y u_y - 2F_z u_z}{c^2}, \frac{F_z u_z - F_y u_y}{c^2}, \frac{F_x u_y + F_y u_x}{c^2}, \frac{F_y u_z + F_z u_y}{c^2}, \frac{F_x u_z + F_z u_x}{c^2}, 0, 0, 0\right)^{\text{T}} \quad (54)$$

The Boussinesq approximation is adopted in the liquid phase domain. Therefore, the external force $\mathbf{F}$ here is the buoyancy force, which is given as:

$$F_i = \rho_{\text{ref}} |\mathbf{g}| \beta (T - T_{\text{ref}}) \delta_{iz} \quad (55)$$

where $\rho_{\text{ref}}$ is the reference density at the reference temperature $T_{\text{ref}}$. $\mathbf{g}$ is the gravitational acceleration. Moreover, $\beta$ is the volumetric expansion coefficient.

The macroscopic physical quantities can be calculated as:

$$\rho = \sum_{\alpha} f_{\alpha} \qquad \rho_0 \mathbf{u} = \sum_{\alpha} f_{\alpha} \mathbf{e}_{\alpha} + \frac{1}{2}\delta_t \mathbf{F} \tag{56}$$

The continuity equation and the momentum equation recovered from this model can be respectively written as:

$$\frac{\partial \rho}{\partial t} + \frac{\partial}{\partial x_j}(\rho u_j) = 0 \tag{57}$$

$$\frac{\partial(\rho u_i)}{\partial t} + \frac{\partial(\rho u_i u_j)}{\partial x_j} = -\frac{\partial p}{\partial x_i} + \frac{\partial}{\partial x_j}\left[\rho \nu \left(\frac{\partial u_i}{\partial x_j} + \frac{\partial u_j}{\partial x_i} - \frac{\partial u_k}{\partial x_k}\delta_{ij}\right)\right] + \frac{\partial}{\partial x_i}\left[\rho \zeta \left(\frac{\partial u_j}{\partial x_j}\right)\right] \tag{58}$$

where the kinetic viscosity and the bulk viscosity are respectively defined as:

$$\nu = \frac{1}{3}\left(\frac{1}{s_\nu} - \frac{1}{2}\right)\delta_t c^2 \qquad \zeta = \frac{2}{9}\left(\frac{1}{s_e} - \frac{1}{2}\right)\delta_t c^2 \tag{59}$$

## 3. Numerical examples

3.1 1 D melting and solidification dominated by conduction

In order to validate the present SRT and MRT models, we will compare our results in three axis directions ($x$, $y$, $z$) of the Cartesian coordinate with the analytical solutions in this section. 1 D melting and solidification problems which have the analytical solutions are selected as the targets in the Section 3.1.1 and 3.1.2, respectively. If one dimension is at $x$ direction, the periodic boundary conditions will be implemented at $y$ and $z$ directions.

3.1.1 melting in a half–space with the initial temperature equalling the freezing point

In this section, melting in a half–space is considered, which is illustrated in Fig. 1. Initially, the solid in a half-space is at the melting temperature $T_m = 0$. When $t = 0$, the temperature at $x/y/z = 0$ is raised to $T_h = 1$, and maintained for $t > 0$. The analytical solution of this problem can be written as [62]:

$$T(i,t) = \begin{cases} T_h - \dfrac{T_h - T_m}{\operatorname{erf}(k)} \operatorname{erf}\left(\dfrac{i}{2\sqrt{t\chi_1}}\right) & 0 \le i \le X(t) \quad \text{liquid} \\ T_m & i > X(t) \quad \text{solid} \end{cases} \quad (60)$$

where $i$ refers to $x, y, z$. $X(t)$ is location of the phase interface at time $t$, which can be calculated as:

$$X(t) = 2k\sqrt{t\chi_1} \qquad \dfrac{Ste_l}{\exp(k^2)\operatorname{erf}(k)} = k\sqrt{\pi} \quad (61)$$

The liquid phase Stefan number $Ste_l = C_{p,l}(T_h - T_m)/L$.

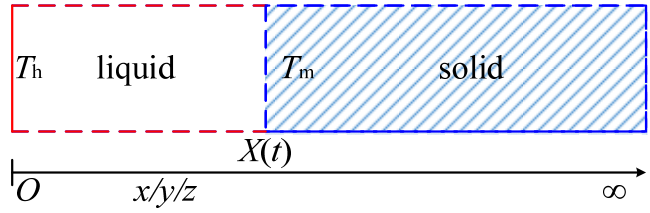

Fig. 2 Schematic of melting in a half–space with the initial temperature equalling the freezing point

In this case, $\chi_l = 0.001$, $C_{p,l} = 1.0$, $Ste_l = 0.1$. Using Eq. (61), we obtain that the location of the phase interface is $x/y/z = 0.44$ when $t = 1000$. As shown in Fig. 3 (b) and (d), both the temperature and the liquid phase fraction distributions at $t = 1000$ calculated by the MRT model agree well with the analytical solutions in $x$ (or $y/z$) direction (the results in $x$, $y$, $z$ direction are uniform) when the relaxation time $\tau = $ 0.564, 1.012 and 6.9, respectively. In another word, this present MRT model is competent to calculate the solid–liquid phase change when the relaxation time $\tau$ varies from 0.564 to 6.9 (or more than 6.9). On the other hand, the SRT model suffers some problems. As Fig. 3 (c) and (d) shows, though the results of SRT model is in good agreement with the analytical solution with $\tau = 1.012$, the temperature and the liquid phase fraction distributions calculated by the SRT deviate from the analytical solution when $\tau = $ 0.564 and 6.9. For $\tau = 0.564$, the temperature at solid phase is -0.00152 rather than the correct $T_m = 0$. In addition, the deviation for $\tau = 6.9$ appears at the phase change interface, which is more obvious. The reason can be found in Section 2.4, which has been mentioned in Ref. [55]. When $\tau = 1.012$ ($\approx 1$), Eq. (47) is approximate satisfied. Therefore, the deviation term is eliminated. However, $\tau = 0.564$ or 6.9 (<<

1 or >> 1) makes the deviation term appear at the phase change interface and disseminate. According to the above analysis, one advantage of the MRT model rises to the surface: the relaxation times in the moment space are not uniform, which makes the model flexible in the selection of each relaxation time. This advantage is useful when appropriate relationship of relaxation times can eliminate the error term.

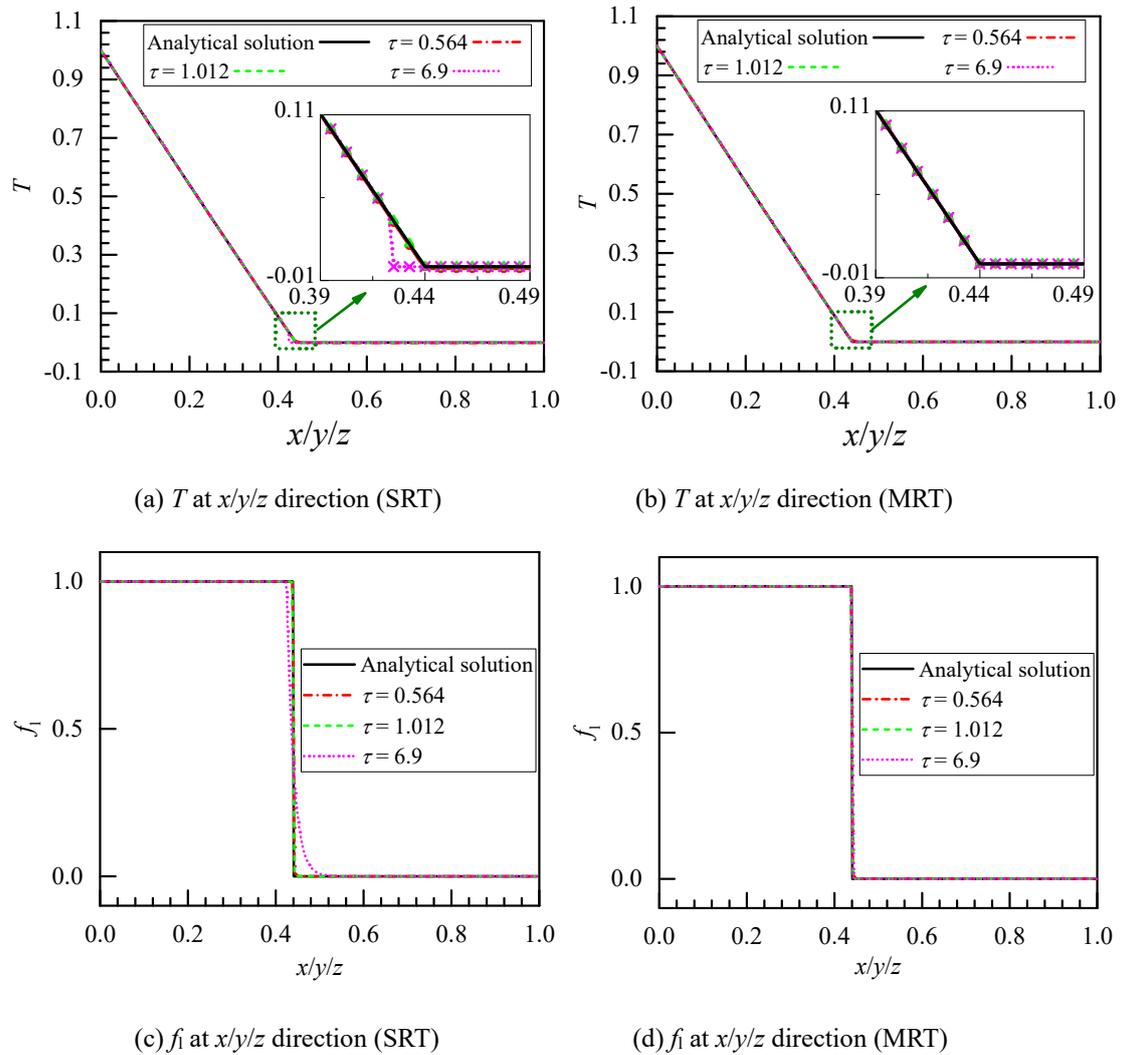

(a) $T$ at $x/y/z$ direction (SRT)  (b) $T$ at $x/y/z$ direction (MRT)

(c) $f_l$ at $x/y/z$ direction (SRT)  (d) $f_l$ at $x/y/z$ direction (MRT)

Fig. 3 SRT versus MRT

Considering the SRT model is practicable only when the relaxation time equal 1 (or close to 1), we never discuss this model in the rest of this paper. In addition, because the results in $x$, $y$, and $z$ direction equal to each other, only the results in $x$ direction are shown in Section 3.1.2.

3.1.2 Solidification in a half–space with the initial temperature more than the freezing point

The initial temperature of the case in Section 3.1.1 equals the melting temperature. More generally, the initial temperature in this section is more than the freezing point. The computational domain is also illustrated in Fig. 1. The initial temperature of the half–space is $T_0 = 1$, while the melting point $T_m$ equals 0. The temperature at $x = 0$ maintains at $T_1 = -1$ when $t > 0$. The analytical solution of this problem can be expressed as [62]:

$$T(i,t) = \begin{cases} T_1 + \dfrac{(T_m - T_1)\mathrm{erf}\left(i/\left(2\sqrt{\chi_s t}\right)\right)}{\mathrm{erf}(k)} & 0 \leq i \leq X(t) \quad \text{solid} \\ T_0 + \dfrac{(T_m - T_0)\mathrm{erfc}\left(i/\left(2\sqrt{\chi_l t}\right)\right)}{\mathrm{erfc}\left(k\sqrt{\chi_s/\chi_l}\right)} & i > X(t) \quad \text{liquid} \end{cases} \qquad (62)$$

where $i$ refers to $x$, $y$, $z$. The parameter $k$ is calculated as:

$$\frac{Ste_s}{\exp(k^2)\mathrm{erf}(k)} - \frac{Ste_l\sqrt{\chi_l/\chi_s}}{\exp(k^2\chi_s/\chi_l)\mathrm{erfc}\left(k\sqrt{\chi_s/\chi_l}\right)} = k\sqrt{\pi} \qquad X(t) = 2k\sqrt{t\chi_s} \qquad (63)$$

The solid phase Stefan number $Ste_s = C_{p,s}(T_m - T_1)/L$, and the liquid phase one $Ste_l = C_{p,l}(T_0 - T_m)/L$. Two cases are simulated in this section. The parameters of the two cases are as follows: (a) $C_{p,s} = C_{p,l} = 1$, $Ste_s = Ste_l = 0.004$, $\lambda_s = \lambda_l = 0.4$; (b) $C_{p,s} = 1$, $C_{p,l} = 2$, $Ste_s = 0.004$, $Ste_l = 0.008$, $\lambda_s = 0.6$, $\lambda_l = 0.15$.

As shown in Fig. 4 (a), the results of case (a) at $t = 0.5$, 3 and 12 are in good agreement with the analytical solutions. For many phase change materials, the specific heat at constant pressure and the thermal conductivity of the solid phase are different from those of the liquid one. Therefore, case (2) is used to check the ability of the MRT model to deal with phase change materials with different thermophysical parameters in solid phase and liquid phase. As shown in Fig. 4 (b), the temperature distributions at time $t = 1$, 3 and 9 calculated by the MRT model are well consistent with the analytical solutions.

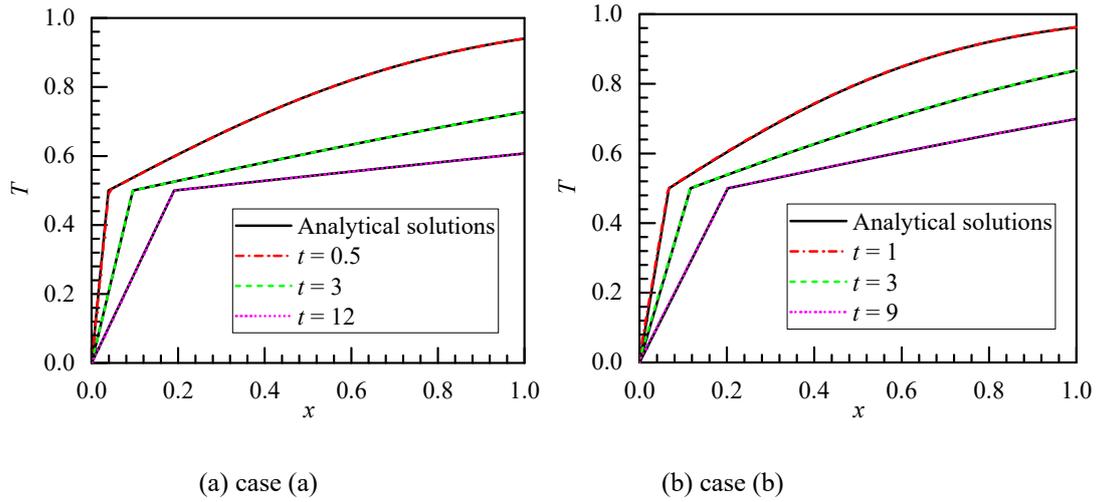

Fig. 4 Temperature distribution at *x* direction

The accurate results near the phase change simulated by the MRT model is benefited from the fact that the recommendatory relationship of the relaxation times mentioned in Section 2.4 eliminates the deviation term generated at the phase interface. In order to vividly illustrate the above explain, we simulate other two cases where the relations of the relaxation times mentioned in Section 2.4 are not satisfied. The physical parameters are same as case (a). The relaxation time $\sigma_1 = \sigma_2 = \sigma_3 = 1.14$ while $\sigma_4 = \sigma_5 = 1.9$ (case (c)) or 0.1 (case (d)) $\neq 2 - \sigma_1$. As shown in Fig. 5, the results with $\sigma_4 = \sigma_5 = 0.1$ deviate more from the analytical solutions than that with $\sigma_4 = \sigma_5 = 1.9$. The relative error of case (c) from the phase interface to $x = 0.21$ is more than 2 % while that of case (a) from the phase interface to the solid phase is about 0.1 %. The error in case (c) seems to transmit from the phase interface to the direction of the movement of the phase interface and decay in the liquid phase domain. On the other hand, the error in case (d) with $\sigma_4 = \sigma_5 = 0.1$ makes the phase interface widen.

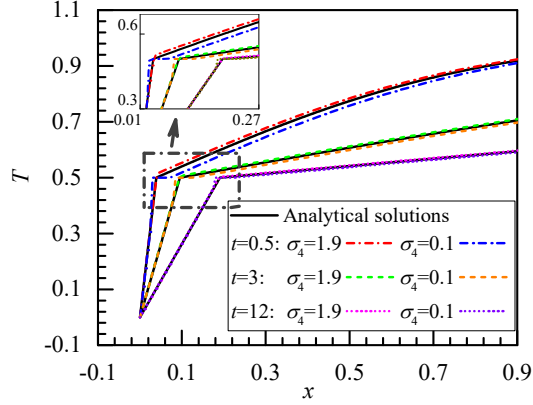

Fig. 5 Temperature distributions at *x* direction of case (c) and case (d)

3.2 2 D melting

After the discussion in Section 3.1, it can be found that the present MRT model has a good performance in each axis (*x*/*y*/*z*). In this section, 2 D cases are used to validate the present model. The *xz* plain is chosen as the computational domain while the periodic boundary condition is implemented at the direction of *y* axis. In Section 3.2.1, 2 D solidification dominated by conduction in a semi–infinite corner is simulated. The present results are compared with both the analytical solutions and other reported numerical results. In Section 3.2.2, the results of 2 D melting with convection calculated by the proposed model will be confirmed by the published numerical results.

3.2.1 2 D solidification dominated by conduction in a semi–infinite corner

Initially, the domain (x > 0 and z > 0) shown in Fig. 6 (a) is liquid phase, and the initial temperature $T_0$ equals 0.3. The temperature at $x = 0$ and $z = 0$ drops to $T_l = -1$ at $t = 0$ and maintains when $t > 0$. Then, solidification begins from the plain $x = 0$ and plain $z = 0$. The freezing temperature $T_f$ is 0. The rest parameters are as follows: $C_{p,s} = C_{p,l} = 1$, $Ste_s = Ste_l = C_{p,l} (T_f - T_l) / L = 4$, $\lambda_s = \lambda_l = 1$. The grid size is 400 × 400. The isotherm distributions and the phase interface (magenta line) at $Fo = 0.25$ calculated by the MRT model are shown in Fig. 6 (b). The quantitative results are presented in Fig. 6 (c) and (d). The present phase interface is reasonably consistent with the analytical result cited in Ref. [63] and the numerical results found in Ref. [53] and [63]. Moreover, the present isotherm distributions are also in good agreement with the numerical results from Ref. [63]and [53].

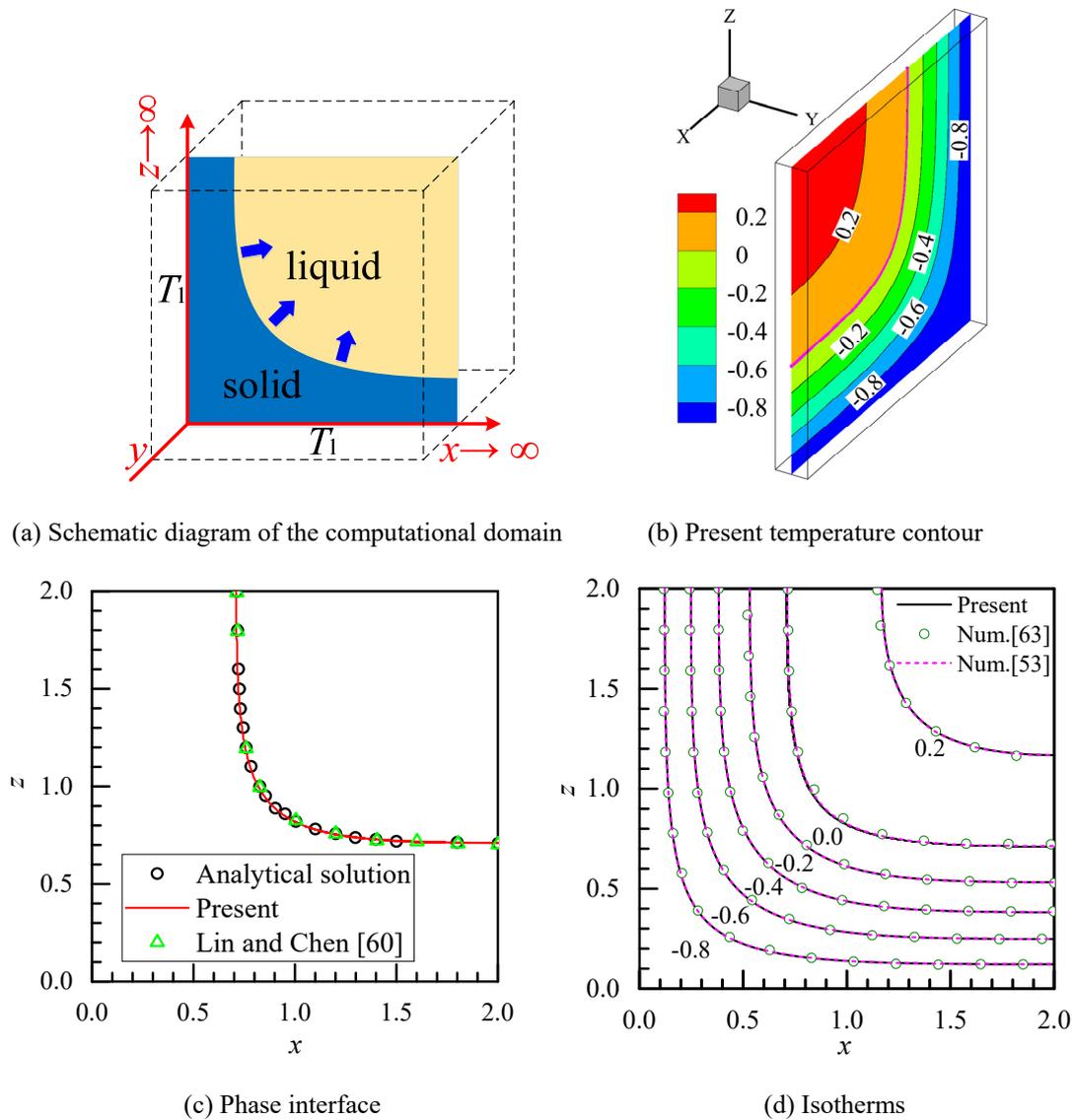

(a) Schematic diagram of the computational domain  (b) Present temperature contour

(c) Phase interface  (d) Isotherms

Fig. 6 Schematic diagram of 2 D solidification in a semi–infinite corner and results at $Fo = 0.25$

3.2.2 2 D melting with convection

2 D solidification without convection in a semi–infinite corner is simulated in Section 3.2.1. More generally, 2 D melting with convection will be discuss in this section. The size of the cuboid cavity (the length at $y$ direction is infinite) illustrated in Fig. 7 is $1 \times 1 \times \infty$. As shown in Fig. 7, the upper (plain at $z = 0$) and lower (plain at $z = 1$) boundaries are adiabatic. Initially, the phase change material filled in the cuboid cavity is solid, whose temperature is uniform and equals the melting temperature $T_m = 0$. The temperature of the plane at $x = 0$ $T_h$ rises to 1 at $t = 0$ and remains unchanged. The temperature of the boundary at $x = 1$ $T_l$ maintains at 0 for times $t \geq 0$. The density, specific heat, and thermal conductivity of the solid phase and those of the liquid phase are equal.

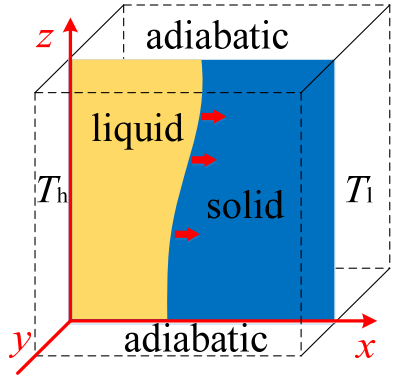

Fig. 7 Schematic diagram of 2 D melting in the cuboid cavity with infinite length at $y$ direction

Four dimensionless parameters used in the phase change with convection, the Rayleigh number $Ra$, the Prandtl number $Pr$, the Stefan number $Ste$, and the Fourier number $Fo$, are defined as:

$$Ra = \frac{|\mathbf{g}|\beta \Delta T l_c^3}{\nu \chi}, \quad Pr = \frac{\nu}{\chi}, \quad Ste = \frac{C_p \Delta T}{L}, \quad Fo = \frac{\chi t}{l_c^2} \qquad (64)$$

where $l_c$ is the characteristic length, $\Delta T$ is the temperature difference between the high temperature boundary and the low temperature one of the natural convection, and $t$ is time. $\Delta T$ equals $T_h - T_m$ and $l_c$ equals 1 in this case. The values of $Ra$, $Pr$, and $Ste$ are listed at Table 1, whose values are as same as those in the case in Ref. [54] for validation.

Table 1 the dimensionless parameters in this case

| $Ra$ | $Pr$ | $Ste$ |
| --- | --- | --- |
| 2.5E4 | 0.02 | 0.01 |

The temperature distributions, streamlines (black lines), and phase interfaces (white line) obtained by the present MRT model and those from Ref. [54] at $Fo = 4, 10, 20$ are both shown in Fig. 8. Initially, owing to the weak strength of the convection, the conduction is dominant in heat transfer, which causes the phase interface is approximately vertical at $Fo = 4$. With time going, the convection becomes more and more strong. Then, the conduction has not the final say. The heat transfer is dominated by both convection and conduction. The phase interface is obviously bent by the convection. Comparing Fig. 8 (a), (c), and (e) with Fig. 8 (b), (d), and (f), we can find that the present results are qualitatively consistent with those obtained by the model in Ref. [54].

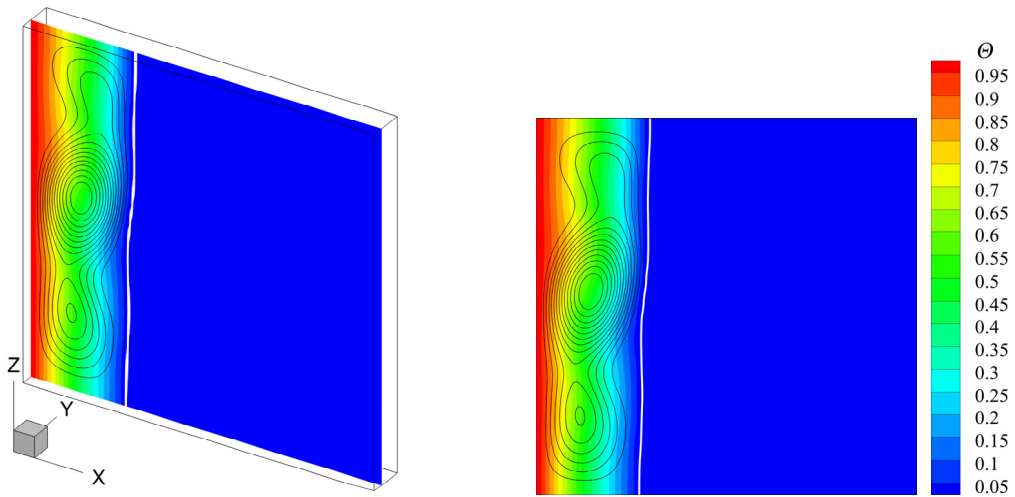

(a) $Fo = 4$ (present)　　　　　　　　　　　(b) $Fo = 4$ (Ref. [54])

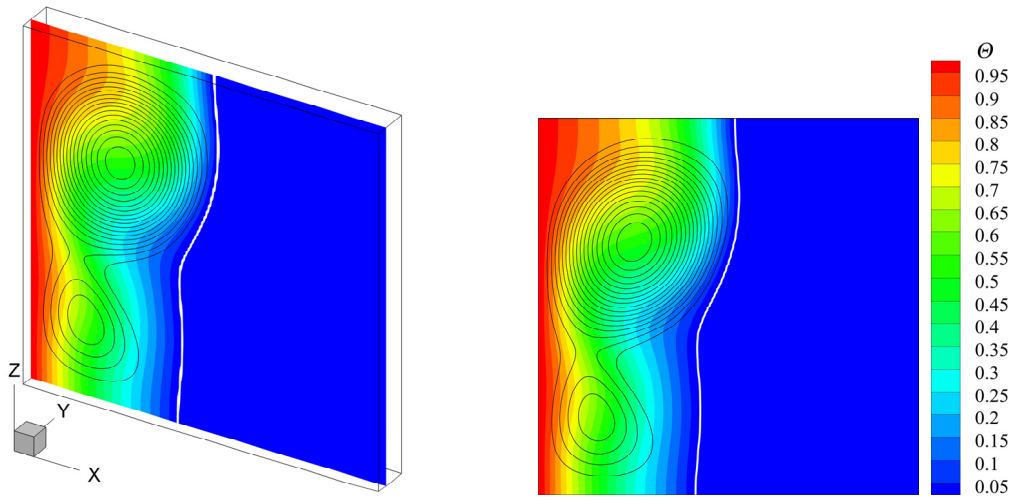

(c) $Fo = 10$ (present)　　　　　　　　　　　(d) $Fo = 10$ (Ref. [54])

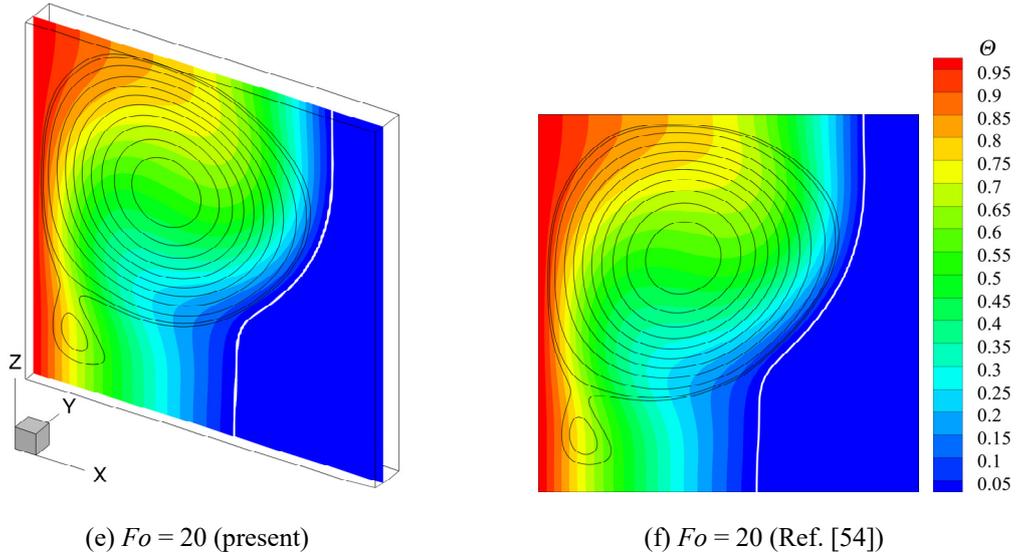

(e) $Fo = 20$ (present)          (f) $Fo = 20$ (Ref. [54])

Fig. 8 the temperature contours, streamlines, and phase interfaces obtained by the present MRT model and reported in Ref. [54]

Furthermore, the quantitative verification is also made. The average Nusselt number $Nu$ along the left wall at the $xz$ plain is defined as:

$$\overline{Nu} = -\frac{1}{l_z}\int_0^{l_z} \frac{1}{(T_h - T_m)}\left(\frac{\partial T}{\partial x}\right)\delta_z \tag{65}$$

Fig. 9 (a) shows the average Nusselt number along the left wall versus Fourier number obtained by the present model, that from Ref. [64], and that reported in Ref. [54]. Moreover, the average liquid phase fraction versus Fourier number is also calculated, as shown in Fig. 9 (b). It can be found that the present results are in reasonable agreement with the results obtained by Mencinger [64] and Huang et al. [54], respectively.

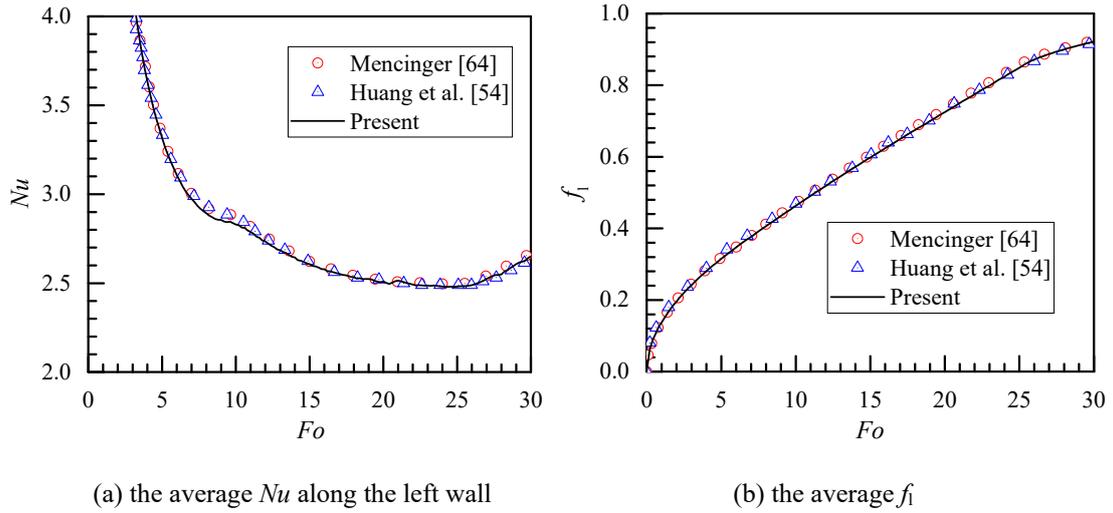

(a) the average $Nu$ along the left wall　　　　(b) the average $f_l$

Fig. 9 the average $Nu$ along the left wall and the average $f_l$ versus $Fo$

3.3 3 D melting with convection

From 1 D (Section 3.1) to 2 D (Section 3.2), seven cases have been calculated by the present MRT model for validation. More generally, 3 D melting will be discussed in this section. The size of the cubic cavity illustrated in Fig. 10 is 1 × 1 ×1. As shown in Fig. 10, the $xy$ plains and the $yz$ plains of the cubic cavity are all adiabatic while the isothermal temperature boundaries are implemented at plain $x = 0$ ($T_h = 1$) and plain $x = 1$ ($T_l = 0$). The no slip boundary conditions are used at the contour of the liquid phase domain: the walls of the cubic cavity and the phase interface facing to the liquid phase domain. The melting temperature $T_m$ equals 0. The thermophysical properties of the liquid phase are same as those of the solid phase. Initially, the phase change material filled in the cubic cavity is solid, whose temperature is uniform and equal to the melting temperature. Three cases are calculated in this section, whose grid sizes and dimensionless parameters are listed in Table 2.

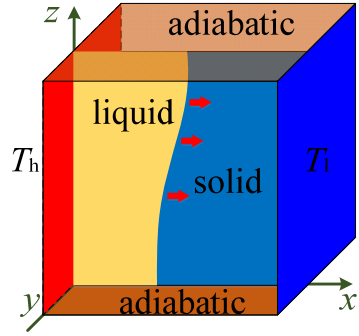

Fig. 10 Schematic diagram of the melting in the cubic cavity

Table 2 parameters in case 1, case 2 and case 3

|  | grid size ($x \times y \times z$) | $Pr$ | $Ra$ | $Ste$ |
| --- | --- | --- | --- | --- |
| case 1 | 160 × 160 × 80 | 0.02 | 2.5E4 | 0.4 |
| case 2 | 160 × 160 × 80 | 10 | 2.5E4 | 0.4 |
| case 3 | 200 × 200 × 80 | 10 | 1.0E6 | 0.4 |

The temperature distributions, streamlines (black lines), and phase interfaces (white lines) of case 1 and case 2 at $Fo$ = 0.2, 0.4, 0.8, as well as those of case 3 at $Fo$ = 0.1, 0.15, 0.4 are shown in Fig. 11. The $Ra$ and $Pr$ of case 1 are same as those of the present 2 D case in Section 3.2.2. However, the number of vortices at $xz$ plain in case 2 is less than that of the 2 D case in Section 3.2.2. The reason may be that unlike 2 D, one dimension more in 3 D makes the flow velocity of the liquid phase is not limited at the $xz$ plain. In another word, the component at $y$ direction of flow weakens the strength of the flow at the $xz$ plain. The similar situation is also found in case 2 and case 3. The value of $Pr$, the ratio of momentum diffusivity to thermal diffusivity, decides whether the flow dominates the heat transfer or not [65]. Therefore, as illustrated in Fig. 11 (a), (b), (c) and (d), (e), (f), the vortices in case 2 are stronger than those in case 1, which causes that the vortices are unsteady in case 1. In case 2, the heat flux entering from the left wall swims with the flow. In addition, quantitative results, shown in Fig. 12, are used to make analysis more authentic. The stronger vortices near the left wall in case 2 result in that the average

*Nu* along the left wall is higher than that of case 1, as Fig. 12 (a) shows. According to the explanation in Ref. [56], if the temperature in the melting zone is not very uniform, higher average *Nu* along the hot wall, owing to higher *Pr*, will lend to higher melting speed, as shown in Fig. 12 (b).

As illustrated in Fig. 11 (d), (e), (f), (g), (h), and (i), the melting speed of case 3 is higher than that of case 2, which is also quantitatively found in Fig. 12 (b). When the *Pr* and the *Ste* is constant, the convection is enhanced with increasing *Ra*, which increases average *Nu* along the hot wall (see Fig. 12 (a)) and accelerates the melting process (see Fig. 12 (b)).

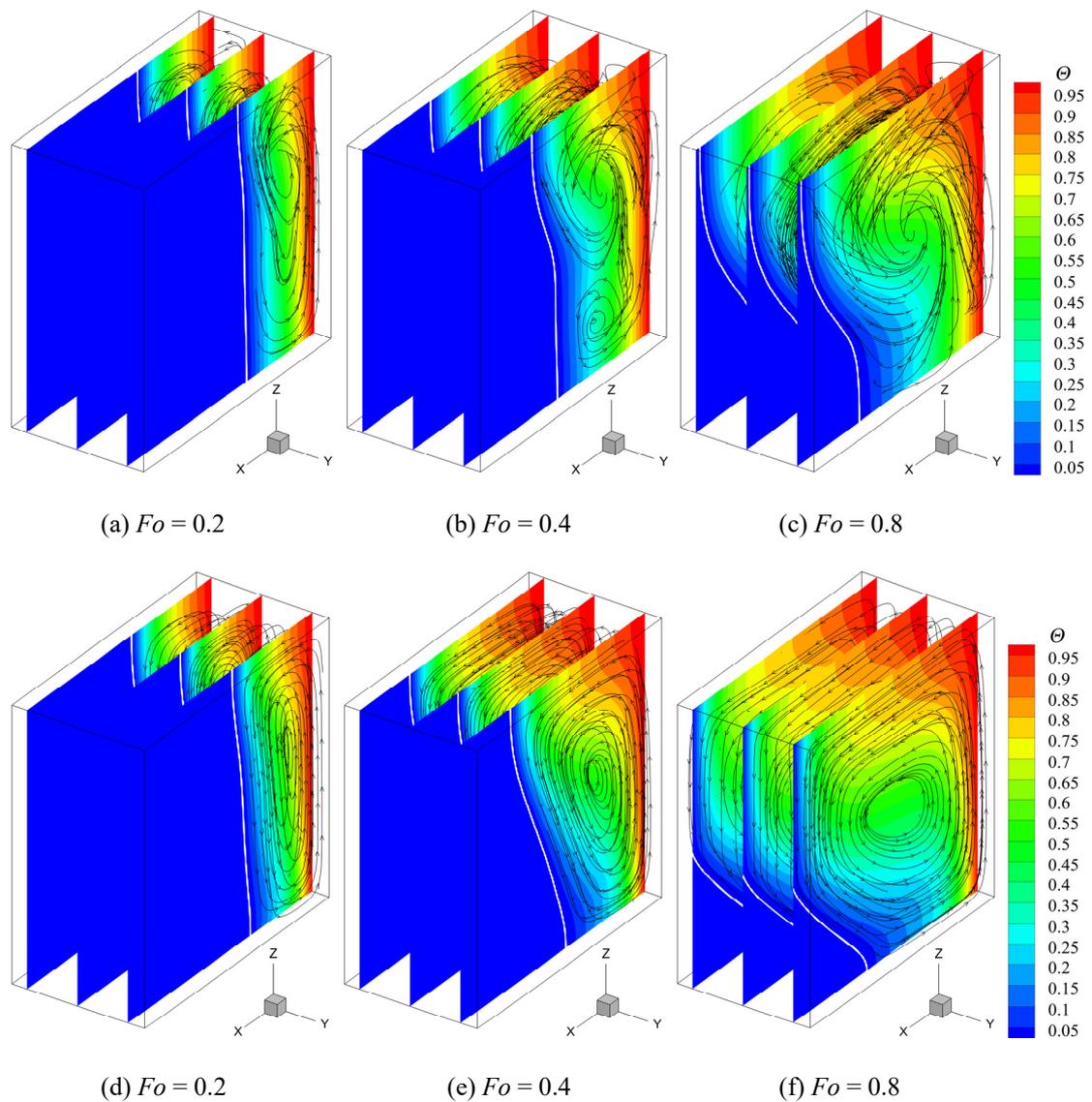

(a) *Fo* = 0.2    (b) *Fo* = 0.4    (c) *Fo* = 0.8

(d) *Fo* = 0.2    (e) *Fo* = 0.4    (f) *Fo* = 0.8

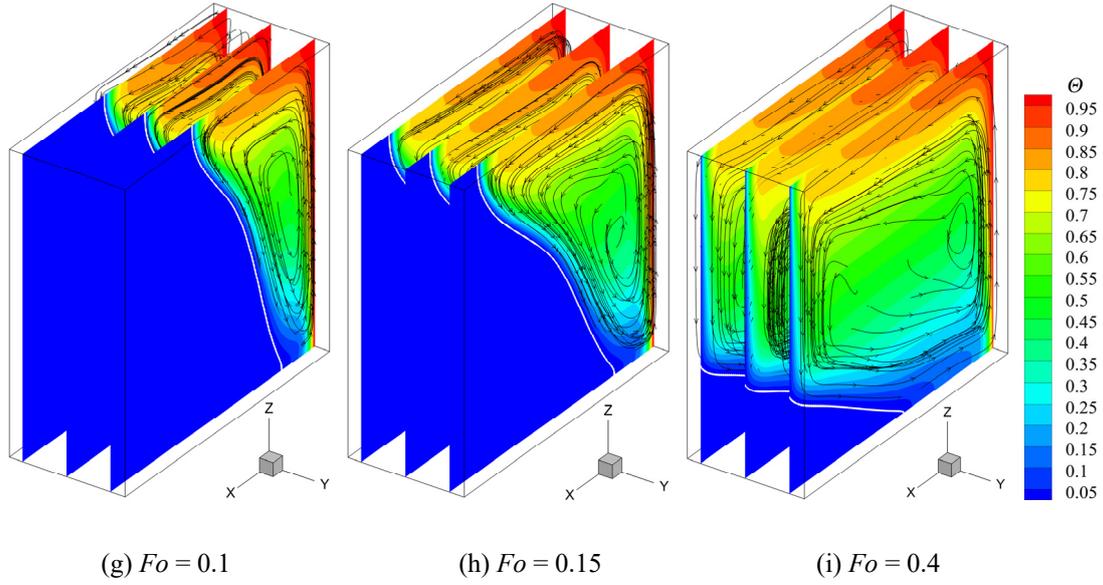

| (g) $Fo = 0.1$ | (h) $Fo = 0.15$ | (i) $Fo = 0.4$ |

Fig. 11 temperature distribution, streamlines, and phase interfaces: case 1: (a), (b), (c); case 2: (d), (e), (f); case (3): (g), (h), (i)

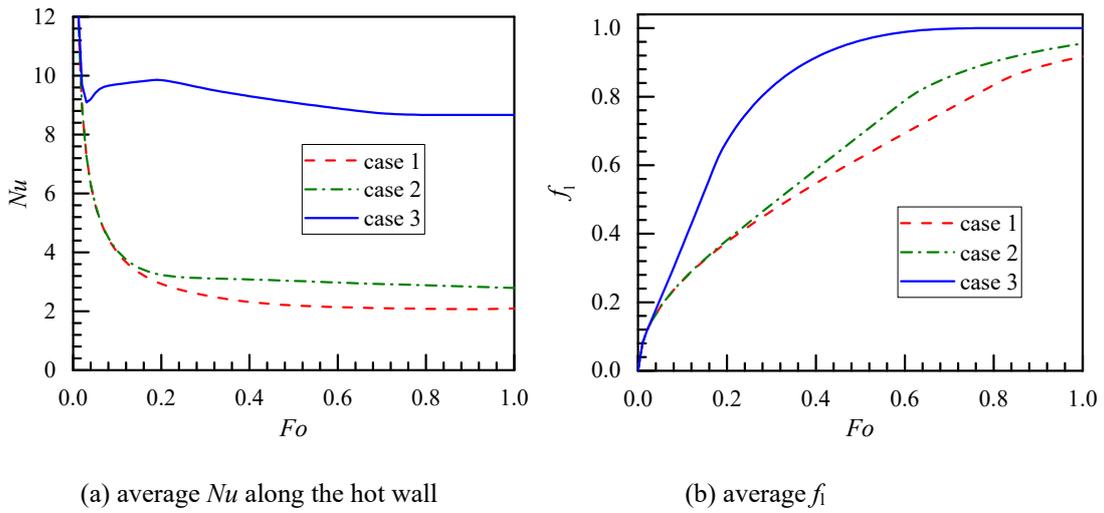

(a) average $Nu$ along the hot wall    (b) average $f_l$

Fig. 12 average $Nu$ along the hot wall and $f_l$ versus $Fo$

## 4. Conclusions

In this paper, both SRT and MRT 3 D enthalpy–based LB models for solid–liquid phase change are proposed. Both 1 D melting and solidification with analytical solutions are respectively calculated by using the SRT and MRT models for validation. Compared with the SRT model, the MRT model is more accurate for the phase interface capture when the relaxation times are not equal to 1, which can be explained by the given analysis of the reasonable relationship of the relaxation times. Moreover, 2 D solidification dominated by conduction in a semi-infinite corner is solved by the present MRT model, which in a good consistency with the analytical solutions and published numerical predictions. In addition, 2 D melting with convection is also simulated. Qualitatively and quantitatively compared with other numerical results, the present results are reasonable and satisfactory. Validations shows that the present MRT model is qualified to solve the 3 D solid–liquid phase change problems. Furthermore, the influences of Rayleigh number and Prandtl number on the 3 D melting are discussed. The melting is accelerated with increasing Rayleigh number. Higher Prandtl number means the more domination of the convection on the heat transfer in melting process.

**Acknowledgments**

This work was supported by the key project of National Science Foundation of China (No.51436007) and the National Key Basic Research Program of China (973 Program) (2013CB228304).